\documentstyle[12pt]{article}
\advance \topmargin by -\headheight
\advance \topmargin by -\headsep

\evensidemargin \oddsidemargin
\begin{document}

\newpage
\setcounter{page}{0}
\pagestyle{empty}
\centerline{{\large\bf Calogero-Sutherland Particles as 
Quasisemions}}
\vskip 50pt
\bigskip
\centerline{Giovanni AMELINO-CAMELIA}
\vskip 18pt
\baselineskip 12pt plus 0.2pt minus 0.2pt
\begin{center}
{\it Center for Theoretical Physics,
MIT,
Cambridge, Massachusetts 02139, USA \\
and\\
Theoretical Physics, University of Oxford,
1 Keble Rd., Oxford OX1 3NP, UK\footnote{Present
address.}}
\end{center}

\vskip 4.5cm

\baselineskip 12pt plus 0.2pt minus 0.2pt
 
\centerline{\bf ABSTRACT }
The ultraviolet structure of the Calogero-Sutherland models
is examined, and, in particular, semions result to have
special properties. An analogy with ultraviolet structures known
in anyon quantum mechanics is drawn, and it is used to
suggest possible physical consequences of the 
observed semionic properties.

\vfill
\noindent{MIT-CTP-2432/OUTP-96-12P \space
hep-th/9609063
\space {\it Modern Physics Letters} A (1996) in press \hfill}

\newpage
\setcounter{page}{1}
\baselineskip 12pt plus 0.2pt minus 0.2pt
\pagestyle{plain}

Recently, there has been renewed 
interest[1-8]
in the Calogero-Sutherland 
models[9-11],
especially in connection with the study of fractional {\it exclusion}
statistics[4,7,12-16]
in 1+1 
dimensions.
These quantum mechanical models describe particles whose dynamics
is governed by a Hamiltonian of the form
\begin{equation}
- \sum_{i}  {d^2 \over dx_i^2} + 
\sum_{i<j} { \pi^2 \beta (\beta -1) \over 
L^2 sin^2 \left( \pi (x_i-x_j)/L \right)} + V(\{x_i\}) ~,
\label{casugen}
\end{equation}
where $i,j \! = \! 1,2,...,N$, $x_i$ denotes the $i$-th particle position
on a circle of radius $L$,
$\beta$ is a non-negative real parameter,
and $V$ is a regular ({\it i.e.} finite for every $\{x_i\}$) potential.
The parameter $\beta$ has been found to characterize
the {\it exclusion} statistics of the particles\cite{pasqlect,bernlect};
in particular, (once appropriate boundary conditions are 
imposed\cite{pasqlect})
$\beta \! = \! 0$ corresponds to bosons
and $\beta \! = \! 1$ corresponds to fermions\footnote{Note that
the potential $sin^{-2}(\pi x/L)$,
which is singular at the points of coincidence
of particle positions and causes the fractionality of the 
{\it exclusion} statistics,
has vanishing coefficient for $\beta = 0,1$.}.

The cases $V \! = \! 0$ (``free Calogero-Sutherland
particles") and $V \! = \! \sum_{i<j} (x_i-x_j)^2$ 
(``Calogero-Sutherland
particles with an harmonic potential") have been 
completely solved\cite{calo2,sutherland} both for finite $L$
and in the limit $L \rightarrow \infty$.

A very important open problem\cite{hanpb,pasqlect,vinet}
is the one of finding a formulation of the Calogero-Sutherland
models in the formalism of non-relativistic quantum field theory.
In the case of anyons\cite{wil}, particles in 2+1 dimensions
that have fractional {\it exchange} statistics\cite{wil},
such a formulation is given by a Chern-Simons field theory, and has 
been very useful[18-20] 
in the understanding of the statistics.

In this Letter I propose a technique of investigation 
of the Calogero-Sutherland
models which should 
allow to uncover some of the features 
of their yet-to-be-found field theoretical formulation.
My analysis is indeed
motivated by an analogy with the case of the anyon models.
In that context 
it has been recently realized\cite{gacbak,lozanob} 
that the ultraviolet structure of
the perturbative expansions in the statistical parameter
is closely related to the structure of the Chern-Simons
field theoretical formulation, which, for example, leads to
Feynman diagrams affected by ultraviolet divergences reproducing
the ones encountered in the quantum mechanical perturbative 
framework\cite{gacbak,lozanob}.
I am therefore interested in an analogous perturbative 
expansion for the Calogero-Sutherland models. 

\noindent
I start by analyzing the ultraviolet problems of such an expansion.
For simplicity, I limit the discussion to the case of two
Calogero-Sutherland particles with $0 \! \le \! \beta \! \le \! 1$,
an harmonic oscillator potential, and $L \! \sim \! \infty$;
the relative motion is therefore described
by the Hamiltonian 
\begin{equation}
H_\beta = - {d^2 \over dx^2} + x^2 + { \beta (\beta -1) \over x^2} ~,
\label{casumy}
\end{equation}
where $x$ is the relative coordinate, and, since we are dealing with
two identical particles on the line, the configuration space 
is $x \! \ge \! 0$.
The harmonic potential is introduced\cite{calo1} in order to
discretize the spectrum, so that the dependence on $\beta$
can be examined more easily.

\noindent
The eigenfunctions of $H_\beta$ that are regular at the point
$x \! = \! 0$, where the particle positions coincide,
are\cite{calo1} (the $L_{n}^{\mu}$ are
Laguerre polynomials and the $N_{n}^{\beta}$ are normalization
constants)
\begin{equation}
|\Psi_{n,\beta}> =
N_{n}^{\beta}~ x^\beta ~
e^{ - {x^2 \over 2}}~
L_{n}^{\beta - 1/2}(x^2)
~, \label{sola}
\end{equation}
and have energies
\begin{equation}
E_{n,\beta} = 4n + 2 + 2 (\beta - 1/2)
~. \label{solb}
\end{equation}

In analogy with the perturbative approaches used in the study of anyons,
one can consider perturbative expansions around zero-th order
solutions $|\Psi_{n,\beta_0}>$,$E_{n,\beta_0}$, 
which would allow to describe the fractional {\it exclusion}
statistics of particles
with any $\beta$ in terms of the one of particles 
with $\beta \! = \! \beta_0$.
Important building blocks of such a  perturbative expansion
are the matrix elements
\begin{eqnarray}
<\Psi_{n,\beta_0}| {1 \over x^2} |\Psi_{n,\beta_0}> \sim 
\int^{\infty}_{0} dx~  {e^{-x^2} [L_{n}^{\beta_0 - 1/2}(x^2)]^2
\over x^{2- 2 \beta_0 }} ~,
\label{eqbf}
\end{eqnarray}
but these are (ultraviolet) divergent for every $\beta_0 \le 1/2$.
An ultraviolet problem
somewhat analogous to this one is encountered
in the study of anyons. In that case
one is
interested in perturbative expansions depending on the statistical 
parameter $\nu$\cite{wil,pap9}, which also has bosonic 
limit $\nu \! = \! 0$ and fermionic
limit $\nu \! = \! 1$,
and one encounters logarithmic ultraviolet divergences
when expanding around the special (bosonic) value $\nu \! = \! 0$.
This divergent bosonic end perturbation theory of anyons can be 
handled\cite{pap9,manu} by using the formalism of
renormalization for quantum mechanics\cite{jakdelta},
and a direct relation between the structure of the renormalized 
perturbative approach and some features of the Chern-Simons
field theoretical formulation of anyons 
has been found[19-21].
The hope that such a program might be completed also for the 
Calogero-Sutherland models 
is confronted by the realization that the ultraviolet
problems illustrated by Eq.(\ref{eqbf}) are much worse than the ones
of the anyon case.
Rather than being specific of a certain choice
of the center of the expansion $\beta_0$, these ultraviolet
problems are encountered for any of a continuous of choices 
of $\beta_0$, and in general the divergences are
worse-than-logarithmic.
However, from Eq.(\ref{eqbf}) one can 
see that the expansion around $\beta_0 \! = \! 1/2$ 
is only affected by logarithmic divergences,
and therefore this type of expansion is the best
candidate for a generalization of the results obtained
for anyons with the bosonic end perturbation theory.

Motivated by this observation,
I now consider more carefully the possibility of studying
Calogero-Sutherland particles with any $\beta$ as perturbations
of ``Calogero-Sutherland semions", {\it i.e.}
Calogero-Sutherland particles with $\beta \! = \! 1/2$.
Following the usual path of renormalization theory in quantum 
mechanics (see, for example, Refs.[20,23-25]),
I add (only for the perturbation theory) the 
counterterm\footnote{As I shall show in a longer paper\cite{gacnext},
the form $\delta(x)/x$ of the counterterm can be derived using 
symmetries and the 
appropriate power-counting rules for renormalization in
quantum mechanics, and the value of the overall 
coefficient $(\beta \! - \! 1/2)$ can be obtained by looking for a 
fixed point of the renormalization group flow.}
$(\beta \! - \! 1/2) \! \sum_{i<j} \delta(x_i \! - \! x_j)/(x_i \! - \! x_j)$
to the original Calogero-Sutherland
Hamiltonian.
A very general verification of the validity of
this procedure will be given in detail
elsewhere\cite{gacnext}, but here I intend to briefly describe
(to second order in the eigenenergies and first order in 
the eigenfunctions)
how the exact two-body solutions (\ref{sola}) and (\ref{solb})
are correctly reproduced in this way.
Let me start by setting up the renormalized quasisemionic
description of the $H_\beta$ eigenproblem.
The zero-th order Hamiltonian, wave functions, and energies
are obviously the ones for semions,
{\it i.e.} $H_{1/2}$, $|\Psi_{n,1/2}>$, and $E_{n,1/2}$ 
[see Eqs.(\ref{casumy}), (\ref{sola}), and (\ref{solb})].
The renormalized perturbative Hamiltonian is
\begin{equation}
H^{Rpert}_{1/2} \equiv 
H_\beta - H_{1/2} + (\beta \! - \! 1/2) {\delta(x) \over x} 
= (\beta \! - \! 1/2) {\delta(x) \over x}
+ { (\beta \! - \! 1/2)^2 \over x^2} ~.
\label{casumypert}
\end{equation}
Moreover, the regularization/renormalization procedure
obviously requires the introduction of a ultraviolet cut-off $\Lambda$,
which is ultimately removed by
taking the limit $\Lambda \rightarrow \infty$.
In quantum mechanics such a cut-off is 
introduced[20,23-25]
in the integrals that characterize
the matrix elements of the ultraviolet-divergent terms of the
perturbative Hamiltonian; for the present case I define
\begin{equation}
\int^{\infty}_{0} dx ~ 
{1 \over x^2}~f(x) \equiv \int^{\infty}_{1/ \Lambda}  dx ~ 
{1 \over x^2}~f(x) 
~,
\label{rega}
\end{equation}
\begin{equation}
\int^{\infty}_{0} dx ~ {\delta(x) \over x} ~ f(x) \equiv
\Lambda f(1/ \Lambda) ~.
\label{regb}
\end{equation}

Since $H_\beta - H_{1/2}$ is quadratic in $(\beta - 1/2)$,
for the first order energies one easily finds
\begin{eqnarray}
E^{(1)}_{n,\beta} 
= <\Psi_{n,1/2}| \, {(\beta - 1/2) \over |x|} \delta(x) \, |\Psi_{n,1/2}> 
= 2 (\beta - 1/2)
~,
\label{eonecoppia}
\end{eqnarray}
and from Eq.(\ref{solb}) one sees 
that $E^{(1)}_{n,\beta} = E_{n,\beta} - E_{n,1/2}$, {\it i.e.}
the first order result 
(\ref{eonecoppia}) already reproduces the exact result,
as expected since the latter is linear in $(\beta - 1/2)$.

The first order eigenfunctions are given by
\begin{eqnarray}
|\Psi^{(1)}_{n,\beta}> & = &
\sum_{m(\ne n)} {<\Psi_{m,1/2}| \, {(\beta - 1/2) \over |x|} \delta(x) \,
|\Psi_{n,1/2}> \over E_{n,1/2}-E_{m,1/2} }
|\Psi_{m,1/2}>
\nonumber\\
& = & - {(\beta - 1/2) \over 2\sqrt{\pi }} \sum_{m (\ne n)}
{L_{m}^{0}(x^2) \over m-n} x^{1/2} \, e^{-{x^2 \over 2}}
~,
\label{psionecoppia}
\end{eqnarray}
which, as it 
can be seen using properties of the Laguerre polynomials, is in
agreement with the expansion of Eq.(\ref{sola}) to 
first order in $(\beta \! - \! 1/2)$.

Concerning the second order energies,
I have verified that
\begin{eqnarray}
<\Psi_{n,1/2}| \, {(\beta - 1/2)^2 \over x^2} \,|\Psi_{n,1/2}> 
= - <\Psi_{n,1/2}| \, {(\beta - 1/2) \over |x|} \delta(x) \,
|\Psi^{(1)}_{n,\beta}> + O({1 \over \Lambda}) ~, \label{proetwo}
\end{eqnarray}
when the matrix elements are regularized 
using the prescriptions (\ref{rega}) and (\ref{regb}).
From Eq.(\ref{proetwo}) one immediately sees that,
in agreement with Eq.(\ref{solb}), 
$E^{(2)}_{n,\beta} \! = \! 0$
in the $\Lambda \rightarrow \infty$ limit.

This completes the announced two-body test of my quasisemionic
description of Calogero-Sutherland particles. Actually, it should be 
clear that this test illustrates the general mechanism that leads 
to renormalized (finite) results; indeed, 
in the study of the many-body problem
one simply encounters many identical copies of the same divergence
(which originate from many copies of the same 
two-body ${1 \over x^2}$--type interaction), 
and they obviously require a corresponding number
of copies of the same two-body ${\delta(x) \over x}$--type counterterm.
This procedure of generalization to the $N$-body problem has been
worked out in detail in the anyon case\cite{pap9},
and will be discussed for Calogero-Sutherland models in \cite{gacnext}.

The quasisemionic
description that I have proposed and tested 
in this Letter should be useful in the 
understanding of 1+1-dimensional fractional {\it exclusion} statistics,
like the corresponding perturbative approach to the study of anyons
has been useful in the understanding of (2+1-dimensional)
fractional {\it exchange} statistics.
Indeed, both here and in the anyon case the original divergences
are due to the singular potentials which cause the fractionality
of the statistics, and therefore the regularization/renormalization
procedure encodes information on the nature of these statistical
interactions.
In particular, the knowledge of the structure of the
new renormalized
perturbative approach
should be of help in the ongoing search\cite{hanpb,pasqlect,vinet}
of a field theoretical formulation of the Calogero-Sutherland models,
which, in particular,
should have (like Chern-Simons non-relativistic field thery, 
but unlike the usual non-relativistic field theory scenario)
a non-trivial renormalization-requiring ultraviolet structure
related to the one I encountered in the present quantum mechanical
treatment.

Additional input for the search of a field theoretical formulation
could come from devising a quasibosonic perturbative approach to the
study of Calogero-Sutherland models, which would be the ideal starting
point for a bosonized field theoretical formulation.
Research in this direction is certainly encouraged by my results
for the quasisemionic perturbative approach, but,
as I showed, 
the structure of the divergences
in the bosonic limit is substantially different from the one of
the divergences I regularized/renormalized here.

The special role played by Calogero-Sutherland semions
in my analysis should have deeper physical
roots (probably related to the special properties of particles 
with $\beta=1/2$ pointed out in Refs.\cite{casuwil,pasqhalf})
than the rather formal ones I
noticed here. In particular, the fact that the renormalizability
of the semionic perturbation theory that I considered
arises in complete analogy with the renormalizability
of the bosonic perturbation theory used for anyons,
might suggest that semions play a special role in (1+1-dimensional)
exclusion statistics, just like bosons have a special
role\footnote{The simplest description of particles with fractional
exchange statistics is in terms of bosonic fields\cite{jackpi}, and only
starting from bosons one can obtain, via renormalization group
equations (see, for example, Refs.\cite{gacbak,lozanob}), 
2+1-dimensional particles with any given statistics and any given
value of the parameter\cite{gacbak,bourdeaumanu2} 
characterizing the consistent ({\it i.e.} leading to a
self-adjoint Hamiltonian)
choices of the domain
of the Hamiltonian.}
in (2+1-dimensional) exchange statistics.

From the point of view of mathematical physics it is noteworthy that
one more application of renormalization in quantum mechanics
has been here found.
There are not many such 
applications
and this one should be particularly easy to examine because
the problem is 1+1-dimensional and all exact solutions are known.
In particular, certain comparisons between the exact solutions
and the renormalization-requiring perturbative results
might lead to insight in the
physics behind the general 
regularization/renormalization procedure; 
for example,
since the exact solutions (\ref{sola})-(\ref{solb})
are well-defined at every scale,
my analysis is consistent
with the idea\cite{gacbak,manu} that 
the necessity of a cut-off
is simply an artifact of the
perturbative methods used, and not a relict of some unknown
ultraviolet physics.

Finally, I want to emphasize that I chose to
consider only the
regular Calogero-Sutherland eigenfunctions because they
have a clearer physical interpretation\cite{calo1} and
allow a scale-invariant\footnote{As I shall show in 
a longer paper\cite{gacnext},
the scale invariance of the boundary conditions satisfied by the
regular wave functions at $x \! = \! 0$
is related to the fact that here it was appropriate
to work at a fixed point of the 
renormalization group flow and therefore I obtained results 
which do not involve a renormalization scale.}
analysis\cite{gacnext},
but, based on the experience with 
anyons\cite{gacbak,bourdeaumanu2,mit2},
I expect that 
additional insight into the nature of 1+1-dimensional fractional 
{\it exclusion} statistics might be gained by
looking at the renormalized perturbative expansion of the
Calogero-Sutherland eigenfunctions that are singular at the
points of coincidence of particle positions.

\bigskip
\bigskip
I want to thank D. Sen for a conversation on recent
results for the Calogero-Sutherland models, which contributed
to my increasing interest in this field.
I also happily acknowledge conversations 
with D. Bak, M. Bergeron, R. Jackiw, 
V. Pasquier, and D. Seminara.

\newpage
\renewcommand{\Large}{\normalsize} 

\begin{thebibliography}{99} \small
\addtolength{\itemsep}{-6pt}

\bibitem{polprl} A.P. Polychronakos,
Phys. Rev. Lett. {\bf 69}, 703 (1992); ibid {\bf 70}, 2329 (1993).

\bibitem{fqhecasu} H. Azuma, S. Iso, 
Phys. Lett. {\bf B331}, 107 (1994). 

\bibitem{haprl} Z.N.C. Ha, Phys. Rev. Lett. {\bf 73}, 1574 (1994); 
Erratum-ibid. {\bf 74}, 620 (1995).

\bibitem{pasq1} F. Lesage, V. Pasquier, D. Serban, 
Nucl. Phys. {\bf B435}, 585 (1995). 

\bibitem{hanpb} Z.N.C. Ha, Nucl. Phys. {\bf B435}, 604 (1995). 

\bibitem{pasqlect} V. Pasquier, {\it A Lecture on the Calogero-Sutherland
Models}, Rep. No. SACLAY-SPHT-94-060 (1994).

\bibitem{bernlect} D. Bernard, {\it Some Simple (Integrable) Models of
Fractional Statistics}, in Les Houches Summer School: Fluctuating Geometries
in Statistical Mechanics and Field Theory, France, 2 Aug - 9 Sep 1994. 

\bibitem{vinet} L.Lapointe and L. Vinet, {\it Exact Operator Solution of 
the Calogero-Sutherland Model}, to be submitted to Commun. Math. Phys..

\bibitem{calo1} F. Calogero, J. Math. Phys. {\bf 10}, 2191 (1969).

\bibitem{calo2} F. Calogero, J. Math. Phys. {\bf 10}, 2197 (1969);
{\bf 12}, 418 (1971).

\bibitem{sutherland}
B. Sutherland, J. Math. Phys. {\bf 12}, 246 (1971);
ibid. {\bf 12}, 251 (1971); 
Phys. Rev. {\bf A4}, 2019 (1971); ibid. {\bf A5}, 1372 (1971).

\bibitem{haldane} H.D.M. Haldane, Phys. Rev. Lett. {\bf 67}, 937 (1991).

\bibitem{mush} M. V. N. Murthy and R. Shankar, 
Phys. Rev. Lett. {\bf 72}, 3629 (1994).

\bibitem{wucasu} Y.-S. Wu, Phys. Rev. Lett. {\bf 73}, 922 (1994).

\bibitem{casuwil} C. Nayak and F. Wilczek, 
Phys. Rev. Lett. {\bf 73}, 2740 (1994).

\bibitem{ng} W. Chen, Y.J. Ng, and H. van Dam,
Rep. No. IFP-505-UNC (1995).  

\bibitem{wil}{J. M. Leinaas and J. Myrheim,
Nuovo Cimento {\bf B37}, 1 (1977);
F. Wilczek,  Phys. Rev. 
Lett. {\bf 48}, 1144 (1982); ibid. {\bf 49}, 957 (1982).
Also see: F. Wilczek, Fractional Statistics and Anyon
Superconductivity,
(World Scientific, 1990).}

\bibitem{seme} G.W. Semenoff, Phys. Rev. Lett. {\bf 61}, 517 (1988).

\bibitem{jackpi} R. Jackiw and S.-Y. Pi, Phys. Rev. {\bf D42}, 3500 (1990).

\bibitem{gacbak} G. Amelino-Camelia and D. Bak, 
Phys. Lett. {\bf B343}, 231 (1995).

\bibitem{lozanob}{G. Lozano, Phys. Lett. {\bf B283}, 70 (1992);
M.A. Valle Basagoiti, Phys. Lett. {\bf B306}, 307 (1993);
D.Z. Freedman, G. Lozano, and 
N. Rius, Phys. Rev. {\bf D49}, 1054 (1994);
O. Bergman and G. Lozano, Ann. Phys. {\bf 229}, 416 (1994).}

\bibitem{wu}{Y. S. Wu, Phys. Rev. Lett. {\bf 53}, 111 (1984); C. Chou, 
Phys. Rev. {\bf D44}, 2533 (1991);
Erratum-ibid. {\bf D45}, 1433 (1992).}

\bibitem{pap9}{G. Amelino-Camelia, Phys. Lett. {\bf B326}, 282 (1994).}

\bibitem{manu}{C. Manuel and R. Tarrach, Phys. Lett. {\bf B328}, 113 (1994).}

\bibitem{jakdelta} {R. Jackiw, in {\it M.A.B. B\`eg memorial volume}, A. Ali
and P. Hoodbhoy eds. (World Scientific,  Singapore, 1991).}

\bibitem{gacnext} G. Amelino-Camelia, in preparation.

\bibitem{bourdeaumanu2}{C. Manuel and R. Tarrach, Phys.
Lett. {\bf B268}, 222 (1991); M. Bourdeau and R.D. Sorkin, 
Phys. Rev. {\bf D45}, 687 (1992).} 

\bibitem{mit2} G. Amelino-Camelia and L. Hua, Phys. Rev. Lett. {\bf 69},
2875 (1992).

\bibitem{pasqhalf} D. Bernard, V. Pasquier, and D. Serban,
Nucl. Phys. {\bf B428}, 612 (1994);
M.R. Zirnbauer, F.D.M. Haldane, Rep. No. cond-mat/9504108.

\end{thebibliography}
\end{document}